\documentclass[twocolumn]{aastex63}
\usepackage{amsmath}

\shorttitle{Distance of SGR 1935+2154}
\shortauthors{Zhong et al. 2020}

\begin{document}
\title{On the Distance of SGR 1935+2154 Associated with FRB 200428 and Hosted in SNR G57.2+0.8}
\author[0000-0002-1766-6947]{Shu-Qing Zhong}
\affil{School of Astronomy and Space Science, Nanjing University, Nanjing 210093, China; dzg@nju.edu.cn}
\affil{Key laboratory of Modern Astronomy and Astrophysics (Nanjing University), Ministry of Education, Nanjing 210093, China}
\author[0000-0002-7835-8585]{Zi-Gao Dai}
\affil{School of Astronomy and Space Science, Nanjing University, Nanjing 210093, China; dzg@nju.edu.cn}
\affil{Key laboratory of Modern Astronomy and Astrophysics (Nanjing University), Ministry of Education, Nanjing 210093, China}
\author[0000-0001-6863-5369]{Hai-Ming Zhang}
\affil{School of Astronomy and Space Science, Nanjing University, Nanjing 210093, China; dzg@nju.edu.cn}
\affil{Key laboratory of Modern Astronomy and Astrophysics (Nanjing University), Ministry of Education, Nanjing 210093, China}
\author[0000-0003-0471-365X]{Can-Min Deng}
\affil{Department of Astronomy, School of Physical Sciences, University of Science and Technology of China, Hefei, Anhui 230026, China}
\affil{CAS Key Laboratory for Research in Galaxies and Cosmology, Department of Astronomy, University of Science and Technology of China, Hefei 230026, Anhui, China}

\begin{abstract}
Owing to the detection of an extremely bright fast radio burst (FRB) 200428 associated with a hard X-ray counterpart from the magnetar soft gamma-ray repeater (SGR) 1935+2154, the distance of SGR 1935+2154 potentially hosted in the supernova remnant (SNR) G57.2+0.8 can be revisited. Under the assumption that the SGR and the SNR are physically related, in this Letter, by investigating the dispersion measure (DM) of the FRB contributed by the foreground medium of our Galaxy and the local environments and combining with other observational constraints, we find that the distance of SGR 1935+2154 turns out to be $9.0\pm2.5\,$kpc and the SNR radius falls into $10$ to $18\,$pc since the local DM contribution is as low as $0-18\,$pc cm$^{-3}$. These results are basically consistent with the previous studies. In addition, an estimate for the Faraday rotation measure of the SGR and SNR is also carried out.
\end{abstract}

\keywords{Magnetars (992); Soft gamma-ray repeaters (1471); Radio transient sources (2008)}

\section{Introduction}
\label{sec:intro}
Very recently, an extremely bright millisecond-timescale radio burst from
the Galactic magnetar SGR 1935+2154 was reported
by \cite{chime20} and \cite{boch20}. More excitingly,
its associated X-ray burst counterpart was also detected by Insight-HXMT \citep{zhang20b,zhang20c,zhang20d,lick20},
AGILE \citep{tav20}, INTEGRAL \citep{mere20}, and Konus-Wind \citep{rid20} telescopes.
Additionally, a subsequent highly polarised transient pulsating radio burst
was detected by the FAST radio telescope with Faraday rotation measure (RM)
+112.3 rad m$^{-2}$ \citep{zhangcf20},
consistent with RM$=+116\pm6$ rad m$^{-2}$ of FRB 200428 \citep{chime20}.
From the previous investigations about the magnetar SGR 1935+2154,
we know that it has a spin period $P\simeq3.24$ s, a spin-down rate
${\dot{P}}\simeq 1.43\times 10^{-11}\,{\rm s}\,{\rm s}^{-1}$, a surface dipole magnetic field strength $B_{\rm p}\simeq2.2\times10^{14}~{\rm G}$, an age $t\sim3.6$ kyr,
and a spin-down luminosity $L_{\rm sd}\sim 1.7 \times 10^{34}\ \rm{erg~s}^{-1}$ \citep{isr16},
hosted in the Galactic supernova remnant (SNR) G57.2+0.8 with a high probability \citep{gae14}.

In the literature, however, the distance of SNR G57.2+0.8
has a large range and remains highly debated even though
various methods have been used, e.g., the statistical radio surface-brightness-to-diameter relation \citep[$\sim9.1$ kpc,][]{pav13}, the empirical relation between the HI column density $N_{\rm H}$
and the dispersion measure (DM) \citep[$11.7\pm2.8$\,kpc,][]{surn16}, and the local standard of rest (LSR) velocity measure via HI absorption feature \citep[$12.5\pm1.5$\,kpc,][]{koth18},
\citep[$4.5-9.0$\,kpc,][]{rana18}, or via CO gas towards the SNR \citep[$6.6\pm0.7$\,kpc,][]{zhou20}.
For SGR 1935+2154, \cite{koz16} gave an upper limit $<10$ kpc through the scattered correlation between the squares of the radii of the emitting areas and the corresponding black-body temperatures, and \cite{mere20} obtained $2.2-7.1$ kpc through the observation of the bright dust-scattering X-ray ring. Note that the methods tracing the SNR are radio-based only and those tracing the SGR are X-ray-based only.
Due to the position of the SGR at the geometric center of the SNR in a relatively
uncrowded region of the Galactic plane \citep{gae14}, to the distance estimates and
approximate ages inferred for the SGR and the SNR, it is believed that they are likely
physically related \citep{koth18}. Moreover, the relatively small age (3.6 kyr) of the SGR supports that its SNR should be visible \citep{zhou20}.
All these pieces of evidence strongly suggest a likely
association between the SGR and the SNR.

In this Letter, we therefore assume that SGR 1935+2154 is indeed associated with SNR G57.2+0.8
and the SNR has the same age as the SGR, and
then use DM by combining with other observational constraints to estimate the distance of the SGR in Section 2.
Our results are displayed in Section 3. A discussion on RM estimate is arranged in Section 4,
and conclusions are drawn in Section 5.

\section{DM Estimate}
\label{sec:DM}
\cite{chime20} and \cite{boch20} reported that
FRB 200428 has an observed DM$_{\rm obs}=332.7$ pc cm$^{-3}$.
The DM$_{\rm obs}$ is mainly contributed by the foreground interstellar medium (ISM)
in our Galaxy (DM$_{\rm Gal}$), the magnetar wind nebula (DM$_{\rm MWN}$), and the SNR (DM$_{\rm SNR}$), that is,
\begin{equation}
{\rm DM_{obs}=DM_{Gal}+DM_{MWN}+DM_{SNR}},
\label{eq:DM_obs}
\end{equation}
where the foreground DM of our Galaxy is
\begin{equation}
{\rm DM_{Gal}}=\int_0^{D} n_e(l)dl,
\label{eq:DM_Gal}
\end{equation}
related to the distance $D$ of SGR 1935+2154 via the Galactic electron density ($n_e$) distribution NE2001 \citep{cor02,cor03} or YMW16\footnote{Throughout the paper, we adopt the Galatic electron model YMW16 encoded in the pygedm package of Python because this model is believed to give more reliable estimates than NE2001 in general
\citep[see Table 6 of][]{yao17}.} \citep{yao17}.

The DM$_{\rm MWN}$ is primarily attributed to O-mode wave and may be given by
\citep[e.g.,][]{yu14,cao17,yang17}
\begin{equation}\begin{aligned}
\mathrm{DM}_{\rm MWN}\simeq 0.082\ \rm{pc}\ \rm{cm}^{-3}\ \mu_{\pm,4}^{2/3}B_{\rm p,14}^{4/3}P_{0}^{-11/3},
\label{eq:DM_MWN}
\end{aligned}\end{equation}
where $\mu_{\pm}=10^4\mu_{\pm,4}$ is the multiplicity parameter of the electron-positron
pairs, $B_{\rm p}=10^{14}B_{\rm p,14}$\,G
and $P=10^{0}P_{0}$\,s are the dipole magnetic field and the rotation period of the magnetar, respectively.

In regard to the DM$_{\rm SNR}$, it depends on ambient medium:
constant density ISM or wind environment.
So we consider the DM contribution by the SNR in two different scenarios as follows.

\subsection{Constant ISM}
It is widely accepted that an SNR has three phases
after a supernova (SN) explosion in constant ISM scenario:
(a) the free-expansion phase,
(b) the Sedov-Taylor phase, and (c) the snowplow phase.
Because SNR G57.2+0.8 has possibly reached the end of the Sedov-Taylor phase or entered the snowplow phase due to
the non-detection of X-ray emission \citep{koth18,zhou20}, the DM$_{\rm SNR}$ from the ionized medium (including shocked SN ejecta and shocked swept ambient medium\footnote{We assume the swept ambient medium is fully ionized in order to acquire an upper limit of the DM$_{\rm SNR}$. Meanwhile, we neglect the unshocked ambient medium in the upstream of the shock since it is neutral hydrogen dominated, as done in \cite{piro18}.}), can be estimated by
\begin{equation}
{\rm DM_{SNR}}\simeq\left\{\begin{array}{ll}
34\ {\rm pc\ cm^{-3}}\ t_{2}^{2/5}E_{51}^{1/5}n_{2}^{4/5}, & t<t_{\rm SP} \\
81\ {\rm pc\ cm^{-3}}\ t_{3}^{2/7}E_{51}^{0.225}n_{2}^{0.737}, & t>t_{\rm SP}
\end{array}\right.
\label{eq:DM_SNR_ISM}
\end{equation}
during the Sedov-Taylor and snowplow phases \citep[e.g.,][]{yang17,piro18},
where $t=10^it_i$ yr is the age of the SNR, $E=10^{51} E_{51}$ erg is the energy of the SN explosion, and $n=10^2 n_2$ cm$^{-3}$ is the number density of a uniform ISM, as well as the snowplow time $t_{\rm SP}\simeq3920\ {\rm yr}\ E_{51}^{0.22}n_{2}^{-0.55}$ \citep[e.g.,][]{dra11}.
The corresponding SNR radius can be written by
\citep[e.g.,][]{tay50,sedov59,dra11,yang17}
\begin{equation}
R_{\rm SNR}\simeq\left\{\begin{array}{ll}
0.84\ {\rm pc}\ t_2^{2/5}E_{51}^{1/5}n_2^{-1/5}, & t<t_{\rm SP} \\
2.44\ {\rm pc}\ t_3^{2/7}E_{51}^{0.225}n_2^{-0.263}, & t>t_{\rm SP}
\end{array}\right.
\label{eq:R_SNR_ISM}
\end{equation}
where we have used the Sedov-Taylor radius independent of the mass of the SN ejecta as the SNR radius \citep{yang17}
rather than the blastwave radius depending on the mass of the SN ejecta \citep{piro18}, because the Sedov-Taylor radius can be a good representation of the SNR radius when the SNR has been well past the free-expansion phase.

\subsection{Wind Environment}
In a wind environment, the SNR evolution has two phases:
the early ejecta-dominated phase and the very late wind-dominated phase,
based on \cite{piro18}. During these phases, the DM$_{\rm SNR}$ is calculated by \citep[see Table 2 of][]{piro18}
\begin{equation}
{\rm DM_{SNR}}\simeq\left\{\begin{array}{ll}
13\ {\rm pc\ cm^{-3}}\ \mu_{e}^{-1}t_{2}^{-3/2}E_{51}^{-3/4}M_{1}^{5/4}K_{13}^{1/2}, &
\\ t<t_{\rm ch} \\
0.088\ {\rm pc\ cm^{-3}}\ \mu_{e}^{-1}t_{3}^{-2/3}E_{51}^{-1/3}K_{13}^{4/3},&
\\ t>t_{\rm ch}
\end{array}\right.
\label{eq:DM_SNR_W}
\end{equation}
where $\mu_{e}$ is the mean molecular weight per electron, $M=M_1\times 1M_{\odot}$ is the mass of the SN ejecta,
$K=5.1 \times 10^{13}\ {\rm g~cm}^{-1}\ \dot{M}_{-5} {v}_{6}^{-1}$ (here the mass-loss rate $\dot{M}_{-5}=10^{-5} M_{\odot}\ \rm{yr}^{-1}$ and the wind velocity $v_{6}=v_{w} / 10^{6}\ \rm{cm}\ \rm{s}^{-1}$),
and the characteristic time $t_{\rm{ch}}=1.9 \times 10^{3}\ {\rm yr}\ E_{51}^{-1 / 2} M_{1}^{3 / 2} K_{13}^{-1}$ separating these phases. This characteristic
time corresponds to a radius $R_{\rm{ch}}=16.8\  {\rm pc}\ M_{1} K_{13}^{-1}$.
Please note that the SNR radius deemed as the blastwave radius can be linked
to $R_{\rm ch}$ and $t_{\rm ch}$ through
the analytic functions \citep[see Table 2 of][]{piro18}
\begin{equation}
R_{\rm SNR}\simeq\left\{\begin{array}{ll}
1.79R_{\rm ch}\left(t/t_{\rm{ch}}\right)\left[1+0.33\left(t/t_{\rm{ch}}\right)^{1/2}\right]^{-2}, & \\
t<t_{\rm ch} \\
\left[1.11\left(t/t_{\rm{ch}}\right)-0.11\right]^{2/3}R_{\rm ch}, &
\\ t>t_{\rm ch}.
\end{array}\right.
\label{eq:R_SNR_W}
\end{equation}

\section{DM Results}
\label{sec:results}
A useful observational constraint for SNR G57.2+0.8 is that
it is an almost circular source with an average diameter about $10^\prime$, i.e., radius $\theta_{\rm r}\approx5^\prime.5$ \citep{koth18}, which is relevant to the SNR radius via the distance of SGR 1935+2154
\begin{equation}
D=\frac{R_{\rm SNR}}{\theta_{\rm r}}.
\label{eq:theta_r}
\end{equation}
Likewise, the observational constraints for ${\rm DM_{obs}}$, $t$, $B_{\rm p}$, and $P$ are also known.
Through the calculation of Equation (\ref{eq:DM_MWN}),
we find that the value of ${\rm DM_{MWN}}$ is far smaller than 1 pc cm$^{-3}$
even if $\mu_{\pm}$ is very large like $10^6$,
so we safely ignore this term in Equation (\ref{eq:DM_obs}) for subsequent calculations.

\begin{figure}
\vspace{-5mm}
\includegraphics[width=0.55\textwidth, angle=0]{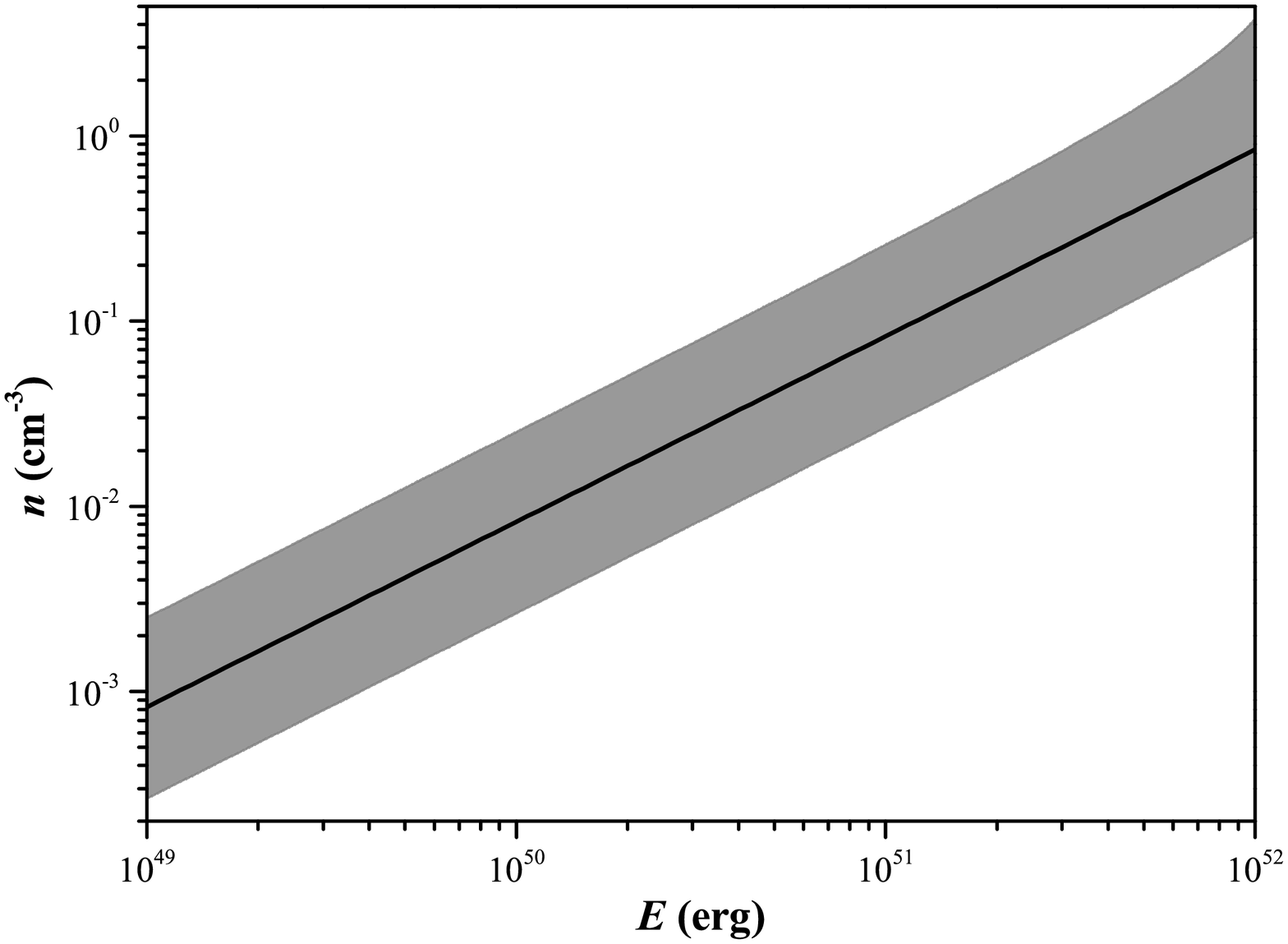}
\includegraphics[width=0.55\textwidth, angle=0]{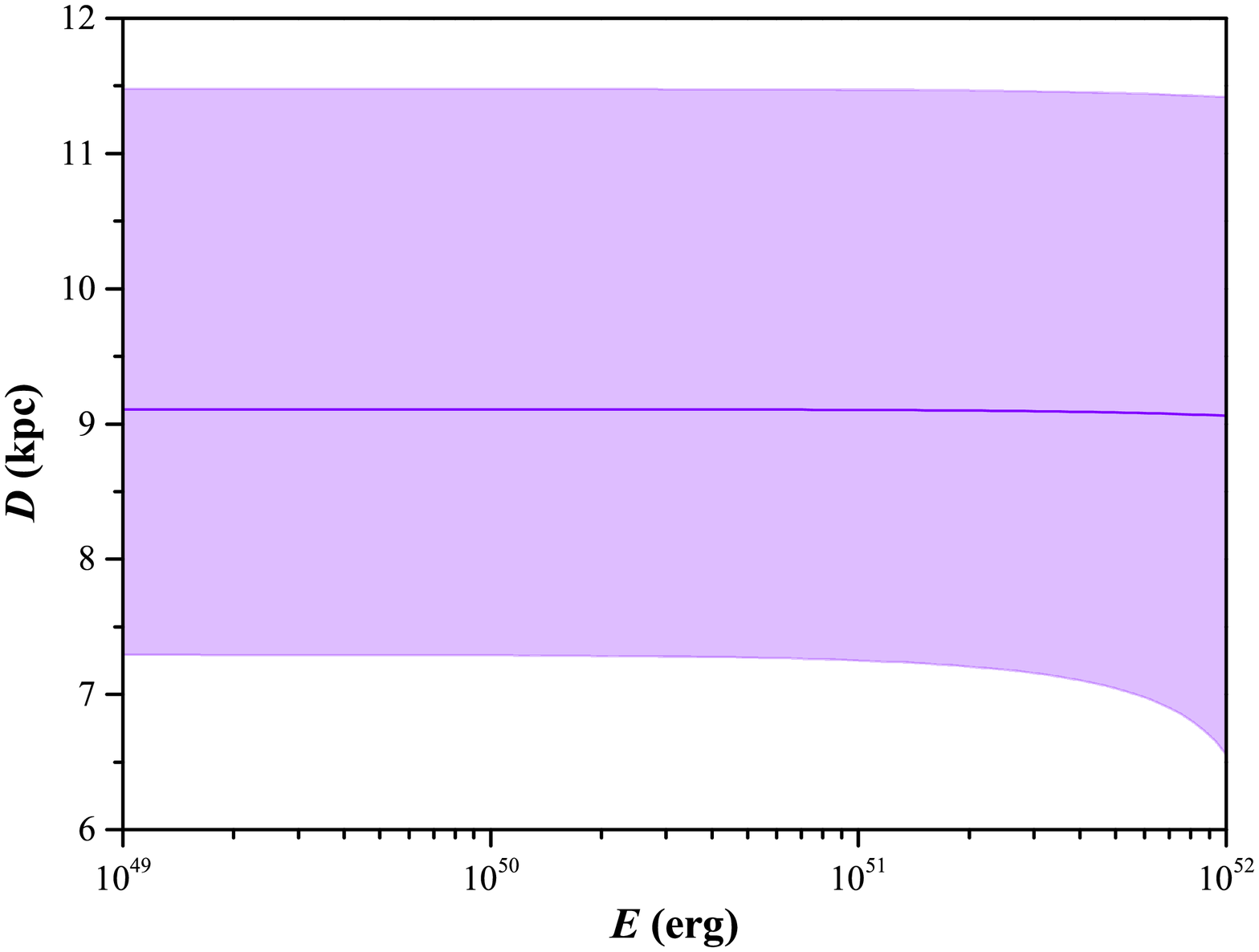}
\includegraphics[width=0.55\textwidth, angle=0]{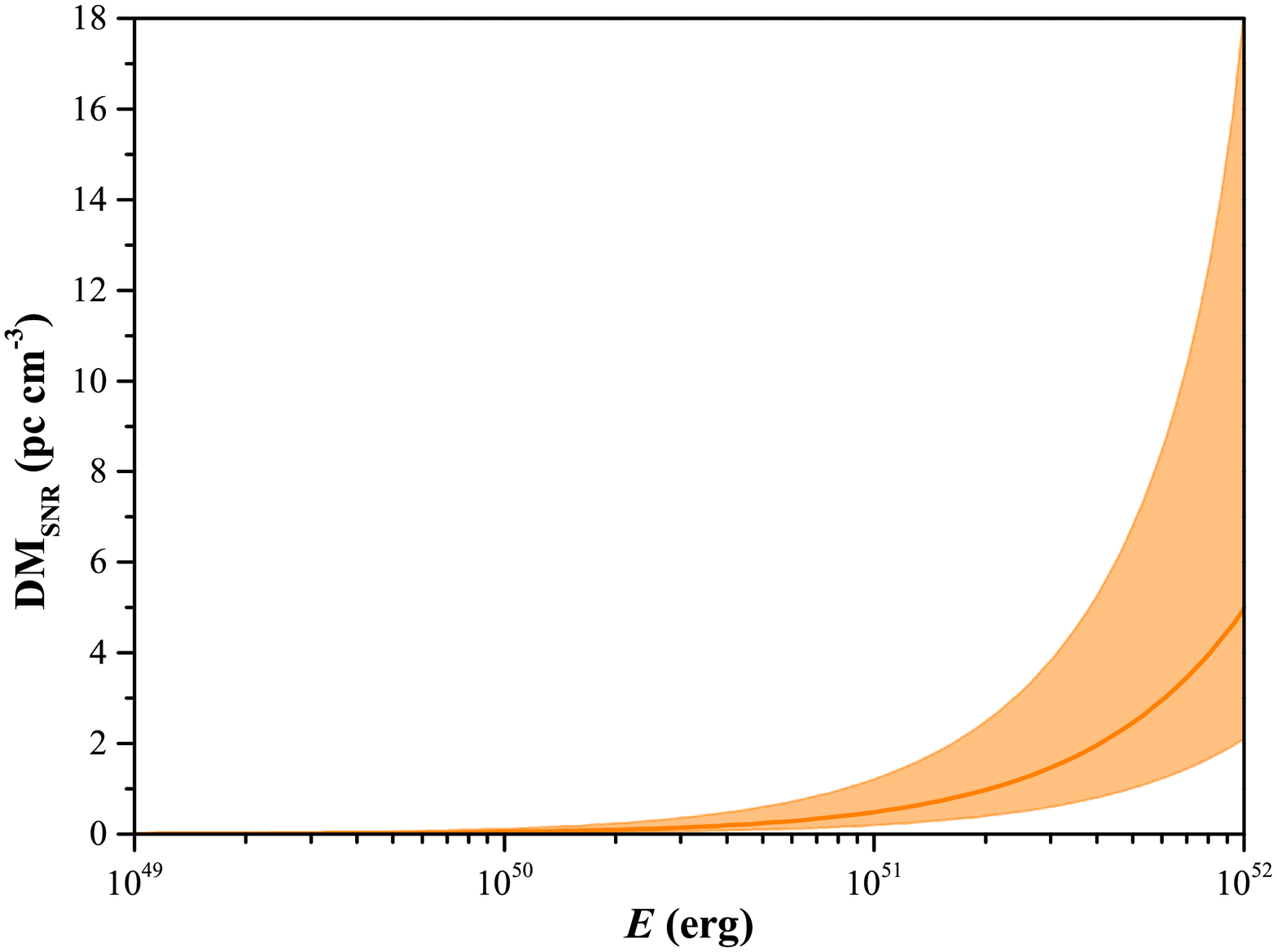}
\caption{In a constant ISM for the SNR: (a) ambient medium density $n$ as a power-law function of energy of SN explosion $E$ (top panel);
(b) the distance $D$ of SGR 1935+2154 is varied with energy of SN explosion $E$ (middle panel);
(c) DM$_{\rm SNR}$ vs. explosion energy $E$ (bottom panel).
The lines in three panels represent the results without considering the uncertainty for the distance estimate via YMW16 model.}
\label{fig:result_ISM}
\end{figure}

In the ISM scenario for the SNR, utilizing Equations (\ref{eq:DM_obs}), (\ref{eq:DM_Gal}), (\ref{eq:DM_SNR_ISM}), (\ref{eq:R_SNR_ISM}), and (\ref{eq:theta_r}),
one gets a power-law relation with an index 1.0
between the explosion energy $E$ and the ambient medium density $n$
(using parameter values $\theta_{\rm r}\approx5^\prime.5$, ${\rm DM_{obs}}=332.7$ pc cm$^{-3}$, and $t=3.6$ kyr),
as illustrated in the top panel of Figure \ref{fig:result_ISM}.
Furthermore, it is obvious that the ambient medium density has a relatively small value, i.e., $<5$ cm$^{-3}$,
within a typical explosion energy ranging from several $10^{49}$ erg
to several $10^{51}$ erg \citep[e.g.,][]{pej15,lyman16}.
Meanwhile, one can also acquire a distance distribution $D\simeq6.5-11.5$ kpc with a mean value 9.0 kpc
(so the SNR radius $R_{\rm SNR}\simeq10-18$ pc),
and a DM distribution of the SNR ${\rm DM_{SNR}}\simeq0-18$ pc cm$^{-3}$ illustrating
in the middle and bottom panels of Figure \ref{fig:result_ISM}.
Obviously, the DM$_{\rm SNR}$ is very low,
compared with the Galactic contribution DM$_{\rm Cal}$.
Note that we have considered the uncertainty for the distance estimate
via YMW16 model throughout the numerical calculations since it is the main uncertainty.
As shown in Table 4 of \cite{yao17}, the direction of SGR 1935+2154 is closest to that of
the pulsar J1932+2220 with a relative uncertainty $D_{\rm err}\sim26\%$ for the distance estimate,
thus the distance of SGR 1935+2154 could also have a relative uncertainty $\sim26\%$.
The lines in Figure \ref{fig:result_ISM} represent the numerical results without considering the uncertainty for the distance.
In reality, it is easy to roughly check these numerical results such as $E\propto n$ via
DM$_{\rm obs}\propto\,\,$DM$_{\rm Gal}\propto D\propto R_{\rm SNR}\propto E^{1/5}n^{-1/5}$ for $t<t_{\rm SP}$
when DM$_{\rm obs}$ is dominated by DM$_{\rm Gal}$.

\begin{figure}
\vspace{-5mm}
\includegraphics[width=0.55\textwidth, angle=0]{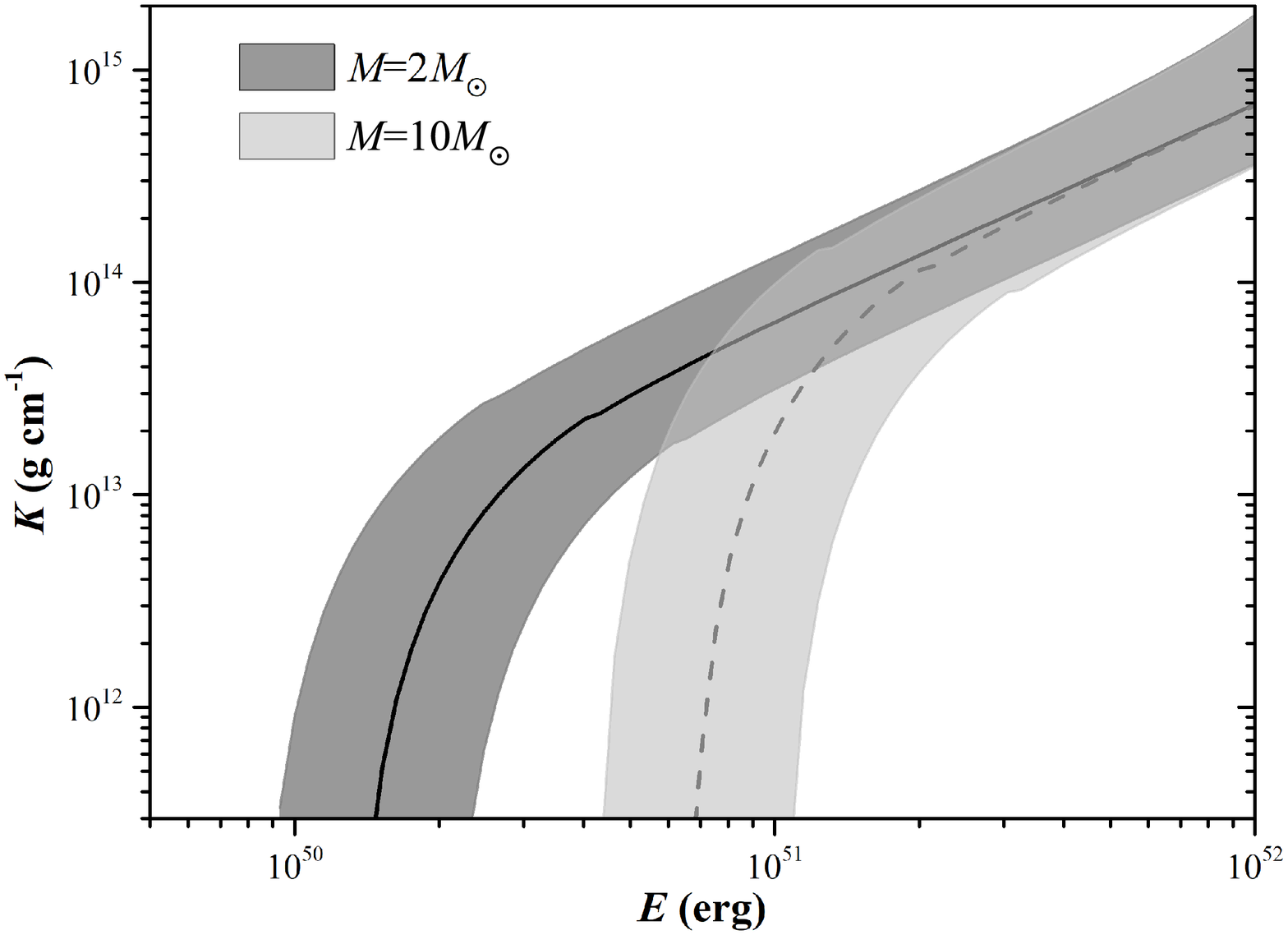}
\includegraphics[width=0.55\textwidth, angle=0]{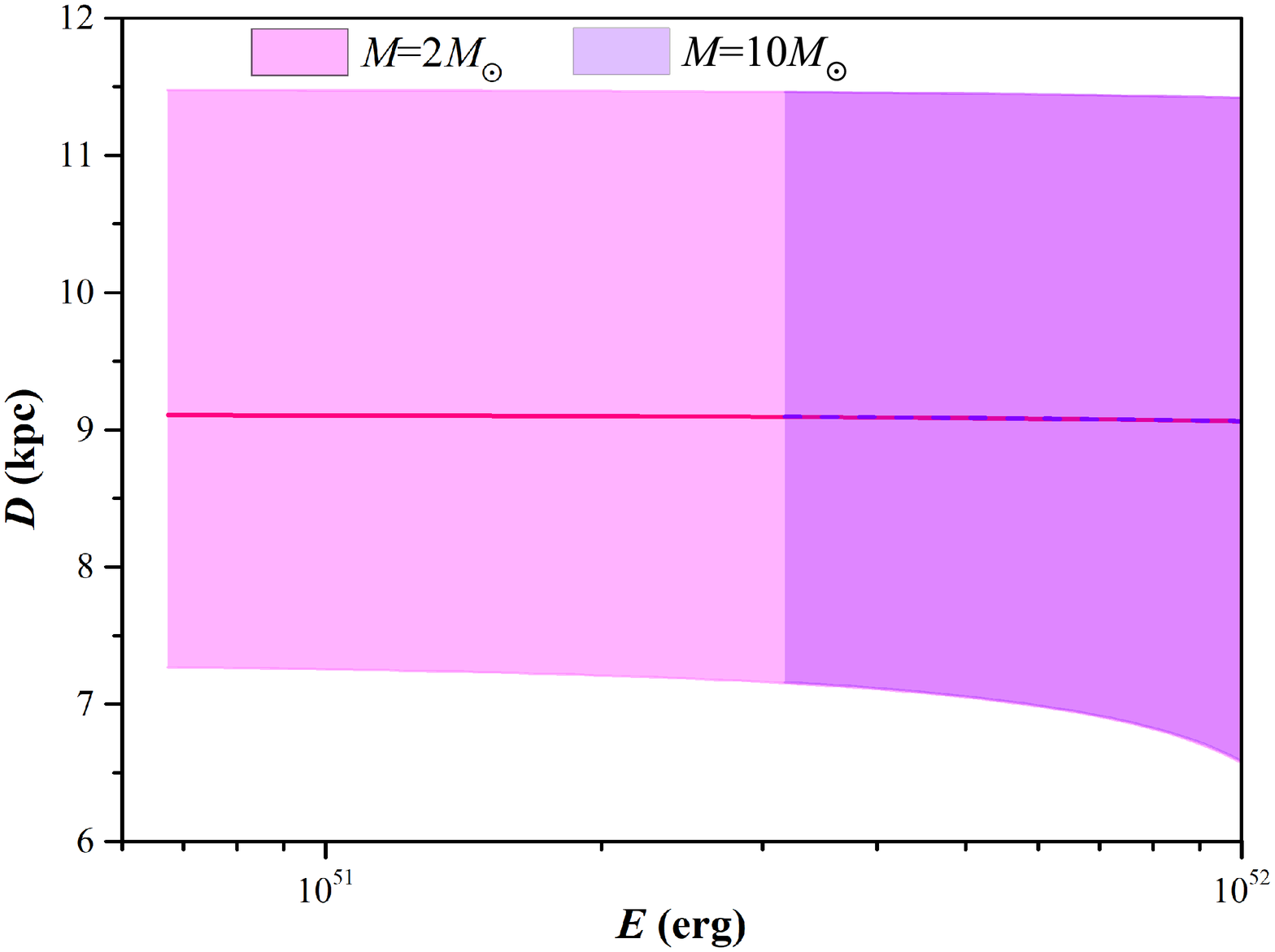}
\includegraphics[width=0.55\textwidth, angle=0]{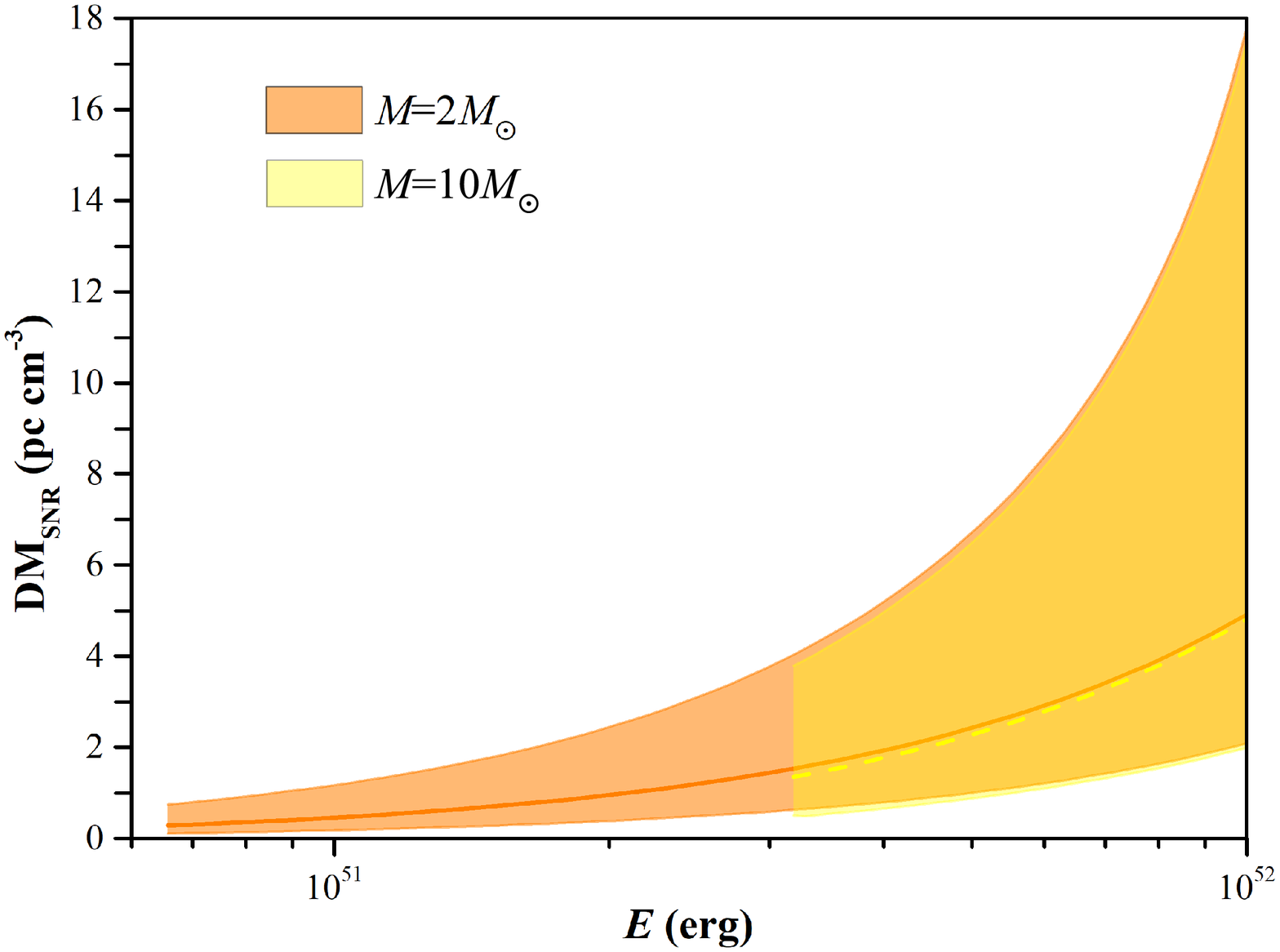}
\caption{In a wind environment for the SNR: (a) the parameter
$K=5.1 \times 10^{13}\ {\rm g~cm}^{-1}\ \dot{M}_{-5} {v}_{6}^{-1}$ as a function of energy of SN explosion $E$ (top panel);
(b) same as the middle panel of Figure \ref{fig:result_ISM} (middle panel);
(c) same as the bottom panel of Figure \ref{fig:result_ISM} (bottom panel).
The lines are also same as Figure \ref{fig:result_ISM}.}
\label{fig:result_W}
\end{figure}

For a wind environment towards the SNR, employing Equations (\ref{eq:DM_obs}), (\ref{eq:DM_Gal}), (\ref{eq:DM_SNR_W})\footnote{Adopting $\mu_{e}=1$. The values of $\mu_{e}$ in a reasonable range cannot significantly influence the final results.},
(\ref{eq:R_SNR_W}), and (\ref{eq:theta_r}), one obtains a relation
between the explosion energy $E$ and the parameter $K$ for $M=2~M_{\odot}$ (stripped-envelope SNe)
and $M=10~M_{\odot}$ (red supergiant progenitors), as shown in the top panel of Figure \ref{fig:result_W}.
The parameter $K$ declines sharply when the explosion energy $E<6\times10^{50}$ erg ($E<3\times10^{51}$ erg)
for $M=2M_{\odot}$ ($M=10M_{\odot}$), so we calculate the numerical results
by only considering the explosion energy $E>6\times10^{50}$ erg ($E>3\times10^{51}$ erg) for $M=2M_{\odot}$ ($M=10M_{\odot}$).
The remaining panels of Figure \ref{fig:result_W} show that the distance spans $D\simeq6.5-11.5$ kpc
and the DM contribution
of the SNR occupies DM$_{\rm SNR}\simeq0-18$ pc cm$^{-3}$ for both $M=2M_{\odot}$ and $M=10M_{\odot}$.
These results are in good agreement with those in the ISM scenario.

In summary, our results generally agree with those in the previous studies
by \cite{pav13}, \cite{surn16},
\cite{koth18}, \cite{rana18},
and \cite{zhou20} for SNR G57.2+0.8,
and \cite{koz16} and \cite{mere20} for SGR 1935+2154. The methods in
\cite{pav13}, \cite{surn16}, and \cite{koz16} are empirical and statistical, with intrinsic large scatter.
Those methods in \cite{koth18}, \cite{rana18},
and \cite{zhou20} seem to be relevant to direct measurements
and their uncertainties mainly stem from the LSR velocity measure and the rotation curve
of the Galaxy. While the uncertainties in the method of \cite{mere20} may result mostly from the determination of the dust layer and the dust-scattering distance. In comparison, the distance estimate from DM in this Letter is assumption-dependent and model-dependent though,
the results are not variable for different ambient environments.
The uncertainty in this method primarily originates from the Galatic electron density distribution
of YMW16 model, i.e., leading to a relative uncertainty $D_{\rm err}\sim26\%$ for the distance in the direction of SGR 1935+2154.

\section{RM Estimate}
\label{sec:RM}
Similar to the DM estimate, the observed RM$_{\rm obs}$ should also have three parts: the foreground RM$_{\rm Gal}$
due to the Galatic ISM and permeating magnetic fields, the RM$_{\rm MWN}$ contributed by the magnetar wind nebula, and the RM$_{\rm SNR}$ resulting from the SNR, that is,
\begin{equation}
{\rm RM_{obs}=RM_{Gal}+RM_{MWN}+RM_{SNR}}.
\label{eq:RM_obs}
\end{equation}

(1) The first part RM$_{\rm Gal}$ can be expressed as
\begin{equation}
\mathrm{RM_{Gal}}[{\rm rad~m^{-2}}]=0.81 \int_{0}^{D} n_{\mathrm{e}}[{\rm cm^{-3}}]B_{\|}[{\rm \mu G}] d l[{\rm pc}]
\label{eq:RM_Gal}
\end{equation}
where $B_{\|}$
is the component of the Galatic magnetic field (GMF) parallel to the line of sight.
RM is positive when the magnetic field points towards us.
There is a general model of the GMF consisting of two
different components: a disk field and a halo field \citep{prou03,sun08}.
The widely used disk field is the logarithmic spiral disk GMF model, which has two versions:
the axisymmetric disk field (ASS model) and the bisymmetric disk field (BSS model)
\citep[e.g.,][]{sima80,han94,stan97,tiny02}.
To estimate the RM$_{\rm Gal}$, we consider the disk field with an ASS or BSS form
and halo field with a basic form \citep{prou03,sun08,jan09,sun10,pshi11}
as done in \cite{lin16}, combining with the Galatic
free electron distribution $n_e$ in \cite{yao17}
and the distance from above DM estimate.
However, the RM$_{\rm Gal}$ has very different values in different models or in same models but with
different parameters,
from a few negative hundred to a few hundred
rad m$^{-2}$ within a distance range of $D\simeq6.5-11.5$ kpc, e.g., $\sim470-750$ rad m$^{-2}$ for ASS+halo
and $\sim50-320$ rad m$^{-2}$ for BSS+halo in \cite{pshi11}, and $\sim-220-40$\,rad m$^{-2}$
for ASS+halo in \cite{sun08}.
As a result, it cannot be well evaluated by the GMF models.
Nevertheless, \cite{koth18} found that the foreground RM$=+223\pm2$ rad m$^{-2}$ for SNR G57.2+0.8
via the polarized intensity maps.

(2) The second part RM$_{\rm MWN}$ arises from
the magnetar wind nebula due to the magnetar spin-down energy release.
The magnetic field of the nebula at time $t$ can be crudely estimated by \citep{metz17}
\begin{equation}
B_{\mathrm{n}} \simeq\left(\frac{6 \epsilon_{B} L_{\mathrm{sd}} t}{R_{\mathrm{n}}^{3}}\right)^{1 / 2},
\label{eq:B_n}
\end{equation}
where $\epsilon_{B}$ is the ratio of the magnetic energy to the shock energy.
Assuming $R_{\rm n}\sim (0.01-0.1)R_{\rm SNR}\simeq 0.1-2$ pc,
and giving $\epsilon_{B}\sim0.1$, $L_{\rm{sd}}\sim1.7\times10^{34}~\rm{erg~s}^{-1}$, and $t\sim3.6$ kyr,
one would get $B_{\rm n}\sim0.5-100\mu{\rm G}$.
In this case, a very low ${\rm{RM_{MWN}}}\simeq0.81\ {\rm rad~m^{-2}}\ \frac{\rm{DM_{MWN}}}{\rm pc~cm^{-3}} \frac{B_{\rm n}}{\mu{\rm G}}\sim0.005-0.3\ {\rm rad~m^{-2}}$ is acquired through Equation (\ref{eq:DM_MWN}).
Although some parameters are uncertain, the RM$_{\rm MWN}$ should be low if they fall into reasonable ranges.

(3) Akin to DM$_{\rm SNR}$ estimate, RM$_{\rm SNR}$ in different surrounding environments should have different
evolutions.

\begin{figure}
\vspace{-3mm}
\includegraphics[width=0.55\textwidth, angle=0]{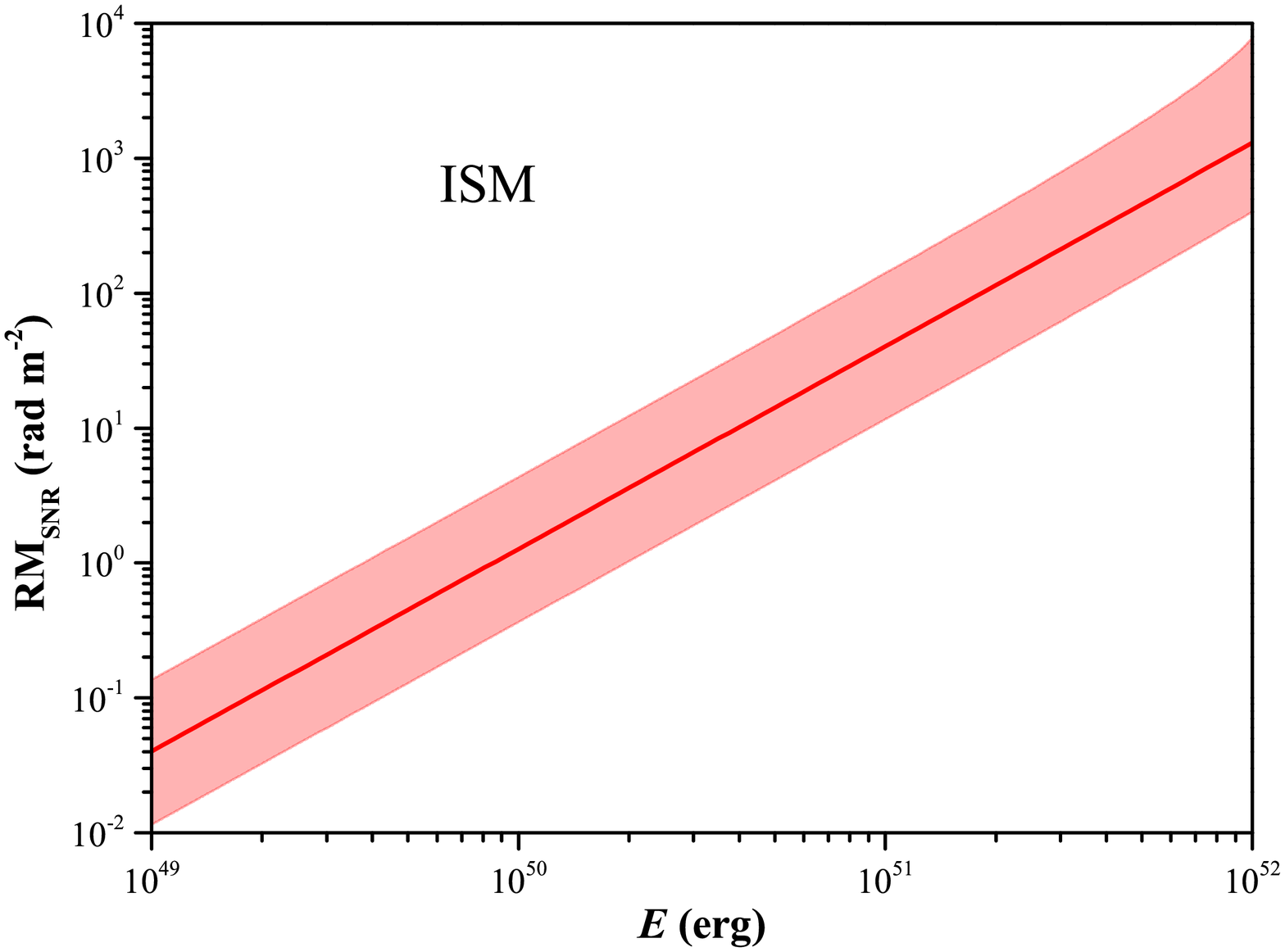}
\includegraphics[width=0.55\textwidth, angle=0]{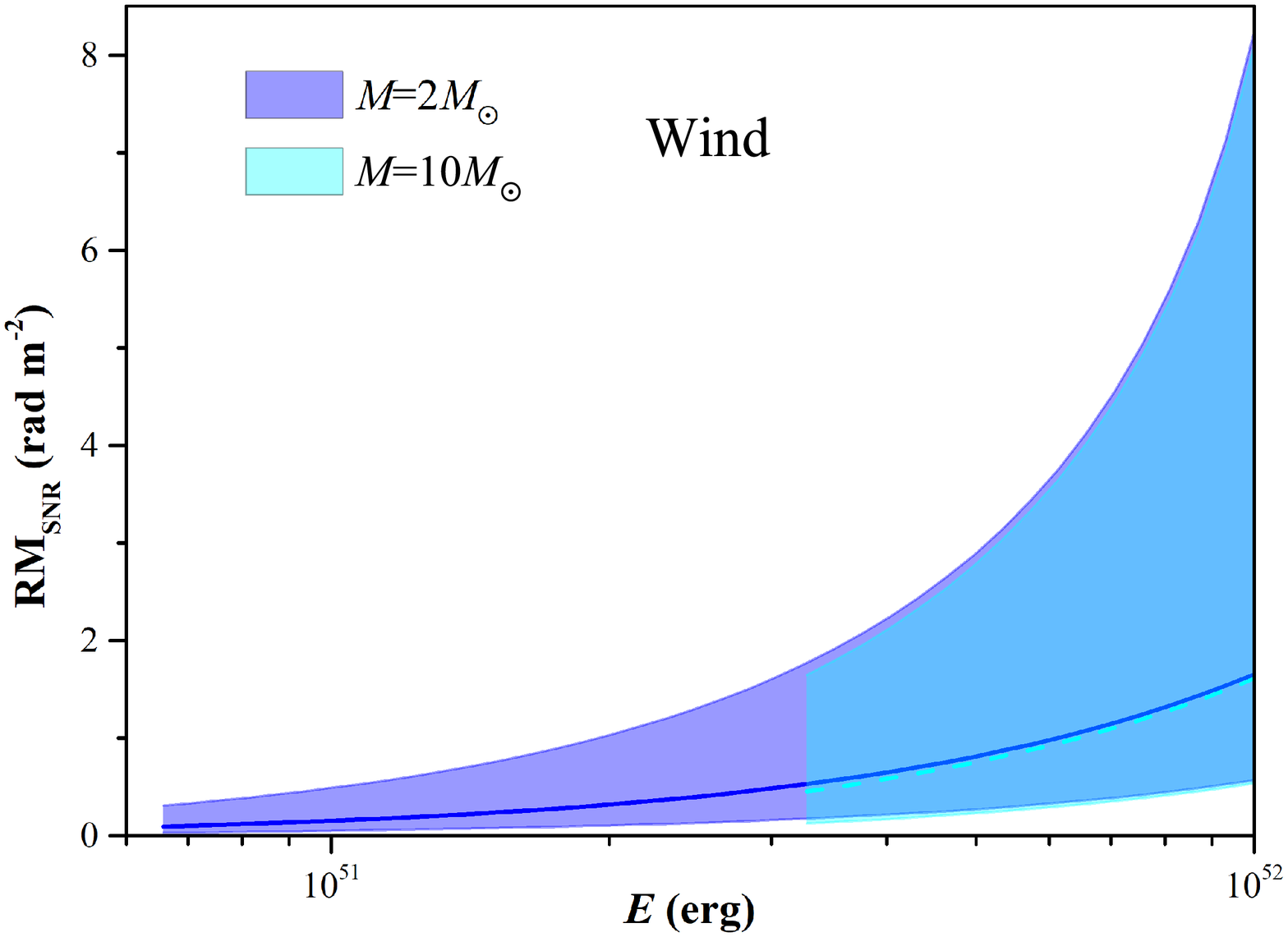}
\caption{RM$_{\rm SNR}$ vs. energy of explosion $E$ in ISM (upper panel)
and wind (lower panel) environments. The lines are same as Figure \ref{fig:result_ISM}.}
\label{fig:RM_SNR}
\end{figure}

{\em ISM Scenario.} In the snowplow phase,
the SNR velocity is \citep{yang17}
\begin{equation}
\begin{aligned} v_{\mathrm{SP}}=690\ \mathrm{km}\ \mathrm{s}^{-1}\
t_3^{-5/7}E_{51}^{0.445}n_2^{-0.813},
\end{aligned}
\label{eq:v_SP}
\end{equation}
so that the magnetic field generated in the shocked ISM is estimated by \citep{piro18}
\begin{eqnarray}
B_{\mathrm{ISM}} &\approx&\left(16 \pi \epsilon m_{p} n\right)^{1 / 2} v_{\rm SP} \nonumber\\
&\approx& 2.02\times10^{3}\ {\rm \mu G}\ \epsilon_{-1}^{1/2}t_3^{-5/7}E_{51}^{0.445}n_2^{-0.313},
\label{eq:B_ISM}
\end{eqnarray}
where $\epsilon=10^{-1}\epsilon_{-1}$ is the ratio of the magnetic energy to the shock energy.
Hence, the RM$_{\rm SNR}$ in the snowplow phase ($t>t_{\rm SP}$) deduced from Equations (\ref{eq:DM_SNR_ISM}) and (\ref{eq:B_ISM}) can be written down as, along with
the RM$_{\rm SNR}$ in the Sedov-Taylor phase ($t<t_{\rm SP}$) \citep[see][]{piro18},
\begin{equation}
{\rm RM_{SNR}}\simeq\left\{\begin{array}{ll}
1.28\times10^5\ {\rm rad~m}^{-2}\ \epsilon_{-1}^{1 / 2} t_{3}^{-1 / 5}E_{51}^{2 / 5} n_{2}^{11 / 10}, &
\\ t<t_{\rm SP} \\
4.94\times10^4\ {\rm rad~m}^{-2}\ \epsilon_{-1}^{1 / 2} t_{4}^{-3/7}E_{51}^{0.67}n_{2}^{0.424},&
\\ t>t_{\rm SP}.
\end{array}\right.
\label{eq:RM_SNR_ISM}
\end{equation}
Combining with the relation between the energy of the SN explosion $E$ and the number density $n$
of ambient ISM in the top panel of
Figure \ref{fig:result_ISM}, one can derive RM$_{\rm SNR}$
as a power-law function of the explosion energy
with an index 1.5,
as displayed in the upper panel of Figure \ref{fig:RM_SNR}. It is also
shown that RM$_{\rm SNR}$ can increase
up to $10^4$ rad m$^{-2}$ when $E$ approaches to $10^{52}$ erg.

{\em Wind Scenario.} The RM$_{\rm SNR}$ in a wind environment is calculated by \citep{piro18}
\begin{equation}
{\rm RM_{SNR}}\simeq\left\{\begin{array}{ll}
0.002\ {\rm rad~m}^{-2}\ x_{0.1}R_{*,2}B_{*,0} \mu_{e}^{-1} E_{51}^{-1} M_{1} t_{3}^{-2}, &
\\ t<t_{\rm ch} \\
0.0017\ {\rm rad~m}^{-2}\ x_{0.1}R_{*,2}B_{*,0} \mu_{e}^{-1} E_{51}^{-2 / 3} K_{13}^{5 / 3} t_{4}^{-4 / 3},&
\\ t>t_{\rm ch}
\end{array}\right.
\label{eq:RM_SNR_W}
\end{equation}
where $x\equiv v_{\rm rot}/v_{\rm w}$ ($v_{\rm rot}$ and $v_{\rm w}$ are the rotation velocity and wind velocity),
$R_{*}=100R_{\odot}R_{*,2}$ and $B_{*}=10^0B_{*,0}$ G are the progenitor's radius and magnetic field, respectively.
Fixing $x=0.1$, $R_{*}=100R_{\odot}$, $\mu_e=1$, and $B_{*}=1$ G (even if they should be variable for
different types of progenitors), and using the relation between the energy of the SN explosion
$E$ and the parameter $K$ in the top panel of Figure \ref{fig:result_W} for different progenitors ($M=2M_{\odot}$
or $M=10M_{\odot}$), one gains a low RM$_{\rm SNR}<8$ rad m$^{-2}$ when the explosion energy $E<10^{52}$ erg,
as exhibited in the lower panel of Figure \ref{fig:RM_SNR}.

Notice that there are a foreground RM$=+223\pm2$ rad m$^{-2}$ for SNR G57.2+0.8 \citep{koth18} and
a RM$=+112.3$ rad m$^{-2}$ for the highly polarised radio burst from SGR 1935+2154 \citep{zhangcf20}.
If this foreground RM has no contribution from the local environment of the SNR,
it would indicate that RM$_{\rm SNR}\sim-110$ rad m$^{-2}$, corresponding to an explosion energy $E\sim(1-5)\times10^{51}$ erg in the ISM scenario from the upper panel of Figure \ref{fig:RM_SNR}.

\section{Conclusions}
\label{sec:conclusions}
In this Letter, we have utilized DMs contributed by the foreground ISM of our Galaxy and the local environments
including the magnetar wind nebula and SNR to estimate the distance of SGR 1935+2154 potentially hosted in SNR G57.2+0.8,
by assuming that the SGR and the SNR are indeed associated and combining with other observational constraints.
Besides, the RM estimate and relevant results have been also discussed.
Some interesting results are summarized as follows:
\begin{itemize}
\item In the constant ISM scenario for the SNR, the energy of the SN explosion $E$ is described by a power-law function as a function of the ambient medium density $n$ with an index 1.0.
    Moreover, the distance, SNR radius, and DM contribution by the SNR are $D\simeq6.5-11.5$ kpc, $R_{\rm SNR}\simeq10-18$ pc, and ${\rm DM_{SNR}}\simeq0-18$ pc cm$^{-3}$ within a typical range of the explosion energy, respectively.
\item In the wind scenario for the SNR, the distance, SNR radius, and DM$_{\rm SNR}$ also spread over similar ranges of those in the ISM scenario for different mass of the SN ejecta.
\item For the RM estimate, the polarization observations from the radio burst of the SGR and the intensity maps of the SNR might signify that the RM contribution by the local environment of the SNR is about $-110$ rad m$^{-2}$ with respect to the explosion energy $\sim(1-5)\times10^{51}$ erg in the ISM scenario for the SNR.
\end{itemize}
Overall, our results relevant to the distance estimate are basically in agreement
with the previous studies.

\acknowledgments
We would like to thank the referee for the very careful and helpful
comments and suggestions that have allowed us to improve the
presentation of this manuscript significantly.
We also thank Wei-Li Lin and Yuan-Pei Yang for their helpful comments and discussions.
This work was supported by the National Key
Research and Development Program of China (grant No.
2017YFA0402600) and the National Natural Science Foundation
of China (grant No. 11833003).
C.M.D. is partially supported by the Fundamental Research Funds for the Central Universities (NO. WK2030000019).


\end{document}